\documentclass[a4paper,aps,prl,twocolumn,floatfix,showpacs]{revtex4}
\usepackage{amsfonts}
\usepackage{amssymb}
\usepackage{graphicx}
\usepackage{texdraw}
\usepackage{color}
\usepackage{subfigure}

\begin{document}

\title{Spontaneous vortex phases in
 ferromagnet-superconductor nanocomposites}
\author{Y. T.  Xing$^{1}$, H. Micklitz$^{1}$}
 \author{T. G. Rappoport$^{2}$}
\author{M. V. Milo\v{s}evi\'{c}$^{3}$}
\author{I. G. Sol\'orzano-Naranjo$^{4}$}
 \author{E. Baggio-Saitovitch$^{1}$}
 \affiliation{$^{1}$Centro Brasileiro de Pesquisas
 F\'isicas, Rio de Janeiro 22290-180,  Brazil}
\affiliation{$^{2}$Instituto de
 F\'{\i}sica, Universidade Federal do Rio de Janeiro, Cx. P. 68528, 219s41-972 , Rio de Janeiro, Brazil}
\affiliation{$^{3}$Departement Fysica, Universiteit Antwerpen,\\
Groenenborgerlaan 171, B-2020 Antwerpen, Belgium}
\affiliation{$^{4}$DCMM, Pontif\' \i cia Universidade Cat\'olica do Rio de Janeiro, Cx. P. 38071, 22453-970,  Rio de Janeiro, Brazil}


\begin{abstract}

We report the appearance of spontaneous vortices in
lead superconducting films with embedded magnetic nanoparticles and
a temperature induced phase transition between different vortex
phases. Unlike common vortices in superconductors, the vortex phase
appears in the absence of  applied magnetic field. The vortices
nucleate exclusively due to the stray field of the magnetic
nanoparticles, which serve the dual role of providing the internal
field and simultaneously acting as pinning centers. Transport
measurements reveal dynamical phase transitions that depend on
temperature (T) and applied field (H) and support the obtained (H-T)
phase diagram.
\end{abstract}

\pacs {74.25.Dw, 74.25.Fy, 74.81.Bd}
\maketitle

The interplay between
of superconductivity (SC) and ferromagnetism (FM) has been attracting
the attention of the scientific community since the discovery of
superconductivity~\cite{AIBuzdin05}. Recently, there has been a
resurgence of this interest due to new phenomena: for example, the
increase of the critical current $J_c$ in hybrid systems containing
sub-micron ferromagnetic particles on top of type II
superconductors~\cite{MLange05,MJVanBael99} and the coexistence of
superconductivity with long-range magnetic order in magnetic
superconductors~\cite{DAoki01}. In a SC-FM hybrid, the magnets strongly affect the properties of the
superconductor, leading to a change in the critical temperature
$T_c$ and critical current $J_c$. Also, they give an opportunity to
observe new phenomena such as domain-wall superconductivity\cite{ZRYang04} and hysteresis pinning effect\cite{APalau07}.

In order to study the characteristics of the novel spontaneous
vortex phases that arise from the interaction between vortices and
embedded magnetic nanoparticles, we fabricated a hybrid system that consists
of a $100$ nm lead film (Pb) containing homogeneously distributed single domain
cobalt (Co) particles (mean diameter about 4.5 nm) with randomly oriented magnetization.

The samples are produced by  the so-called
inert-gas(Ar) aggregation method with an Ar pressure of about
10$^{-1}$ mbar. It is a co-deposition of Pb and well-defined Co
clusters directly onto a sapphire substrate without buffer and
capping layer. The Ar is absorbed by a cryopump at the other end of
the cluster chamber and only well-defined Co clusters can enter
the main chamber. The substrate is mounted on a coldfinger of a
rotatable $^{4}$He cryostat and is cooled to $\sim$ 40 K during the
deposition. The samples are deposited at low temperature in order to get high quality Pb films. The angle between the matrix and the cluster beams was 45$^{o}$. Due to the
different beam directions, samples with different Co volume fraction
can be made within one preparation. The deposition rates are
controlled by three quartz balances in order to monitor the
deposition rate at different positions of the substrate. Ag contacts
for transport measurements are pre-deposited and connected to a
multi-channel automatic measurement system. The typical dimensions
of the sample were 10mm$\times$3mm$\times$100nm. After deposition,
transport properties in both zero and non-zero magnetic field were
investigated in-situ with a built-in split-coil superconducting
magnet (B $\leq$ 1.2T).

This method has some advantages: first, the size of the Co nanoparticles
is tunable and its distribution is very narrow, with a standard deviation of less than 1nm.
 Second, due to the fact that we do not use any buffer or capping layer, any measurement
is related exclusively to the Pb-Co sample. Third, the magnetic moments of the Co particles are randomly oriented so that they have zero total magnetic moment in the sample. After deposition, the samples were annealed at 300 K in order to decrease the defect density. The sample of Co
clusters for the microstructure study was deposited on a carbon foil
which was mounted on a TEM-catcher. A more detailed description of
the experimental set-up and operating procedures for the same equipment
and a similar system of Ag matrix and Co particles can be found in the
literature~\cite{SRubin98}.

\begin{figure}[h]
\includegraphics[width=\columnwidth]{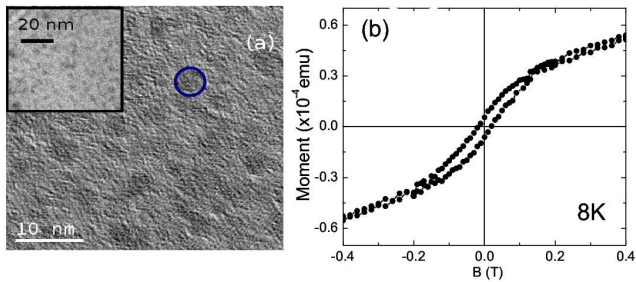}
\caption{(Color online) (a) TEM image of the Co nanoclusters. (b) Hysteresis loop for a sample with 5$\%$
Co volume fraction at 8K. \label{fig1}}
\end{figure}

The morphology of the Co nanoparticles was inferred by transmission
electron microscopy (TEM). As displayed in Fig.~\ref{fig1} (a),
they are very homogeneous in shape with a mean diameter of d$\sim$4.5
nm, therefore smaller than the vortex core size (see below).
Fig.~\ref{fig1}(b) charts a typical magnetic hysteresis loop
measured with a SQUID at a temperature just above $T_c$, indicating
that the Co clusters are indeed ferromagnetic. Similar
nanocomposites have been produced
previously~\cite{APalau07,ASnezhko05,THAlden66}. However, the
appearance of vortices in these systems in the absence of applied
magnetic field has never been observed before. The nanocomposites
were mainly used to investigate the pinning effect of magnetic
nanoparticles in the superconducting phase in the presence of an
external magnetic field.

Theoretical calculations predict that due to their stray field,
ferromagnets inside a superconductor can lead to flux bundles in the
form of spontaneous vortices, antivortices, loops and even closed
loops~\cite{MMDoria07,MMDoria04}. In contrast with other systems in
which superconductivity and ferromagnetism
coexist~\cite{ZRYang04,DAoki01}, the ferromagnetic constituents in
the present case are small single domain particles with randomly
oriented magnetization. Both the size of the particles and the mean
distance between them are smaller than the coherence length ($\xi$) and penetration depth ($\lambda$). The Co clusters have considerable magnetic moments and thus a strong magnetic stray field. If the diameter of the particle (4.5 nm) is much smaller than $\lambda \sim$ 80 nm, the difference of the stray field between a free magnetic particle and a particle in a superconductor is less than one percent. It means the effect of the superconductivity on the stray field of Co particles can be ignored in our case. The maximum field strength of this stray field for a single domain Co particle is about 1.3 T at the poles (corresponding to the Co bulk saturation magnetization of about 1.8 T) and, therefore, much larger than the critical field of $B_c$(0) = 0.1 T for pure Pb. This fact together with the well-known suppression of superconductivity due to the proximity effect caused by non-superconducting metallic particles embedded in a superconductor(eg. Cu particles in Pb~\cite{ISternfeld05})makes a favorable scenario for the formation of spontaneous vortices inside the superconducting Pb film.

Transport measurements give a signature of the presence of vortices in a SC
and thus can be used to characterize a vortex state. In order to
probe the nucleation of the spontaneous vortices in our films, we
analyze the temperature (T) dependence of the resistivity ($\rho$).

\begin{figure}[h]
\includegraphics[width=\columnwidth]{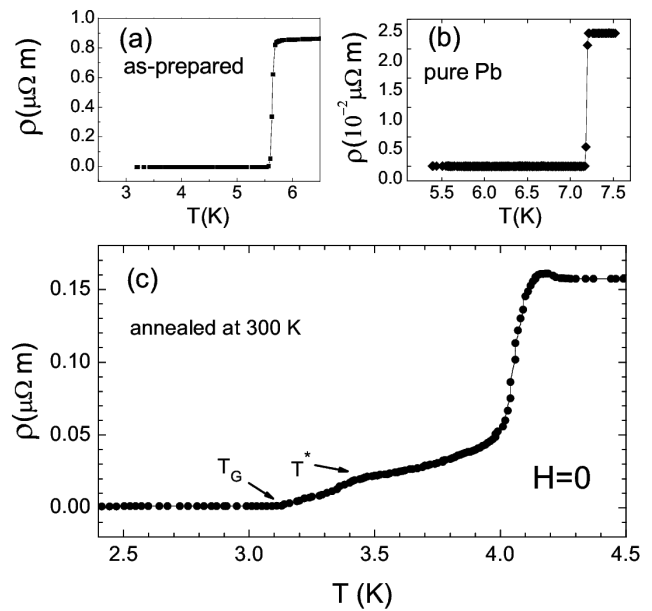}
\caption{Resistivity ($\rho$) versus temperature (T) for (a) an as-prepared film of
lead containing about 3.7 v\% cobalt nanoparticles at zero applied magnetic field, (b) a reference sample of pure lead prepared by the same method and (c) the same sample in (a) but annealed at 300 K. $T_G$ is the temperature that characterizes the transition to a
zero-resistance state. $T^*$ is a temperature that indicates the
kink in the resistivity. The current used in all measurements was $I=0.1$ mA.} \label{fig2}
\end{figure}

In Fig.~\ref{fig2}, the $\rho$-T curves of different samples are shown. Fig.~\ref{fig2} (a) gives the $\rho$-T curves for an as-prepared film (deposition temperature: 40K) with a Co concentration of 3.7 v\%. One can see a sharp transition at a critical temperature $T_c\sim$5.5 K. Compared to $T_c\sim$7.2 K of an undoped pure Pb film, deposited at 40 K and annealed at 300K (Fig.~\ref{fig2} (b)), the Co-doped sample has the expected reduction $T_c$ caused by the two effects discussed above, namely, the proximity effect and the formation of spontaneous vortices. If we compare our $T_c$ reduction ($\Delta T_c \sim$ 1.7 K or $\approx$ 0.5 K/v\%Co) with that found in granular Cu/Pb structures~\cite{ISternfeld05} where the $T_c$ reduction is due to the proximity effect only ($\Delta T_c \sim 0.12$ K/v\%Cu), we can conclude that the effect of spontaneous vortex formation in our Pb/Co system probably is the dominating effect in the $T_c$ reduction. If should be mentioned, however, that this reduction in $T_c$ due to 3.7 v\% Co clusters is much smaller than that caused by Co atoms in Pb~\cite{EWassermann65}. There a decrease of 2 K/v\% Co is observed, which is caused by pair-breaking due to spin flip scattering at paramagnetic impurities. In the case of our Co clusters, having a superparamagnetic blocking temperature of $\sim 25$ K, such a process can not occur below $T_c$.

Annealing the sample at 300 K has dramatic effects(see Fig~\ref{fig2} (c)): it not only reduces the resistivity in the normal state by a factor of $\sim$5, but also changes the superconducting transition. Besides a further reduction of $T_c$, a strongly broadened and structured transition is observed, having zero resistivity at low temperature but an ohmic region between 3.1 and 3.6 K which is followed by a small kink and a second increase of the of the resistivity below the main transition to the normal state at about 4 K. It is important to point out that although the main feature of the $\rho$-T curves for several studied sample are similar, their exact form and the characteristic temperatures depends on the Co volume fraction~\cite{YTXing08}.

In the following we will discuss in detail the structure in the $\rho$-T curve of the annealed sample shown in Fig. ~\ref{fig2}. From the resistivity of the annealed sample we can estimate the superconducting coherence length $\xi$ of this sample. Since the upper critical field $H_{c2}$ in our case is given by a
sum of external applied field and the stray field of the Co
particles, we avoid estimating the coherence length $\xi$ from
$H_{c2}$. Instead we assume the dirty limit for type II SCs and get
the mean-free path from the normal state resistivity (see
Fig.~\ref{fig2} (c)) to determine the coherence length from the
following expression:
\begin{equation}
\xi=0.855(\xi_0l)^{1/2}\sqrt{1-\frac{T}{T_c}}
\end{equation}
where $\xi_0 = 83$ nm is the coherence length of pure lead at zero
temperature and $l$ is the mean-free path. The values of $l$ were
obtained from the resistivity using the free electron model.
The sample mean-free path is $l$ = 3.1 nm leading
to an estimated $\xi\sim25$ nm for temperatures close to 3 K. As
discussed above, the ratio between $\xi$ and $\lambda$ indicates
that the sample is a type II superconductor.  The anomalous $\rho$-T curve for $H=0$ is a first
indication of the existence of more than one phase of vortices or
other forms of trapped flux inside the superconductor.
Close to the critical temperature, we also see a resistance anomaly that
is absent in the non-annealed samples and it is destroyed by
magnetic field. It is known that similar excess resistances were reported in superconducting Al nanostructures~\cite{PSanthanam91,CStrunk98,HVloeberghs92} and inhomogeneous superconducting films~\cite{APark97}. Some possible origins for the phenomena are  normal-superconducting (N-S) interfaces induced  by dynamic phase slip centers~\cite{PSanthanam91,CStrunk98} or non-homogeneous distribution of critical temperatures inside the sample~\cite{HVloeberghs92}. This anomaly was also reported in a system composed of a SC doped with magnetic impurities~\cite{PLindqvist90}. We do not know the exact reason for the appearance of the resistance peak in our sample. However, it is of no relevance for the feature in the $\rho$-T curve far below the onset of the superconductivity.

This low temperature $H=0$ behavior we see in Fig.~\ref{fig2} (c)
is similar to the one observed in type II SCs ,
such as Y-Ba-Cu-O~ containing random point defects~\cite{AMPetrean00} or
amorphous SCs such as MoSi~\cite{NCYeh93} under an applied magnetic field. The curves suggest that
at very low temperatures the vortices are organized in a solid state
phase. Due to the random orientation of the magnetic particles and
presence of loops, we presume that the vortices are highly entangled
leading to a kind of vortex glass. Above 3.1 K, the onset of the
ohmic behavior corresponds to a phase transition with thermally
induced vortex movement. The kink at $T^*$ represents a transition
between liquids in different pinning regimes or the rearrangement
and breaking of the vortex loops without depinning from the pinning
centers. However, from these measurements, it is difficult to
characterize the phases. We could attribute this resistivity
behavior to either a spontaneous creation of vortices that connect
to the sample boundaries, some other type of trapped vortex
state~\cite{MMDoria07} or flux creep~\cite{PWAnderson62}.

\begin{figure}[h]
\includegraphics[width=\columnwidth,clip]{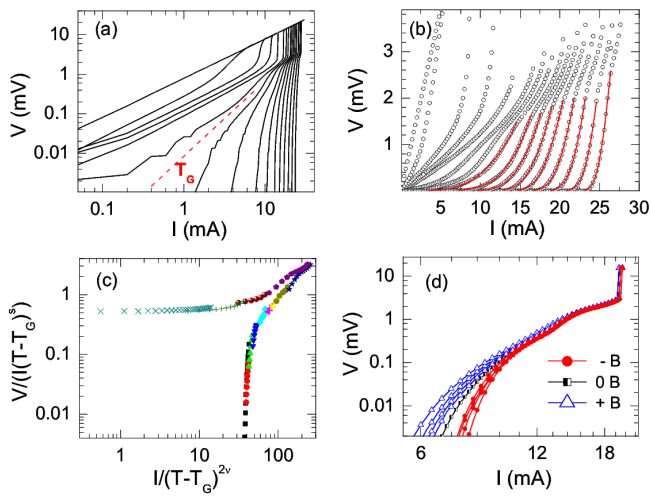}
\caption{(Color online) (a) Isotherms of voltage (V) versus current (I) in a
log-log scale with no magnetic field applied and T are 2.26, 2.35, 2.44, 2.53, 2.6, 2.72, 2.8, 2.9, 3.0, 3.1, 3.2, 3.4, 3.6, 3.8, 4.0 and 4.2 K from right to left. (b) Low-voltage region of the same isotherms of V-I. (c) Scaling analysis for the same data presented
in (a) from T varying from 2.44K to 3.8K. The exponent $s$ is given
by $s=\nu(z-2+D)$. In the curves different colors indicate different temperatures. (d) Effect of positive and negative magnetic field (varying from -0.01 T to 0.01T) in the V-I curves at 2.9 K. For details, see main text.  \label{fig3}}
\end{figure}

The most convincing evidence of the existence of a spontaneous
vortex solid (SVS) at low temperatures comes from the
characterization of the vortex dynamics by means of isothermal
voltage (V)-current (I) measurements. Fig.~\ref{fig3}(a) and
Fig.~\ref{fig3}(b) show a set of (V-I) curves for the same sample of
Fig.~\ref{fig2} at H = 0 and temperatures ranging from 2.26 K to 4.2
K. They indicate the existence of spontaneous vortices networks or
loops induced solely by the magnetic nanoparticles' stray field. For
a better understanding of the relation between the $\rho$-T and
(V-I) curves, we focus on the curves below the main transition
(Fig.~\ref{fig3}(b)): for low temperatures, the voltage is zero for
low currents, which indicates the existence  of a vortex solid.
Furthermore, it increases exponentially with increasing current,
a characteristic of both vortex glass  and flux creep~\cite{GBlatter94}.
For intermediate temperatures, the V-I curves have an S-shape,
 associated to vortex movements with two
different activation energies. Finally, it increases monotonically
with current for temperatures just below $T_c$, as it is expected
for a normal vortex liquid.

On a logarithmic scale (Fig.~\ref{fig3}(a)) a simpler picture
appears, with a clear separation of two classes of (V-I) curves: one
for low temperatures, below a critical line $T_G$ and the other for
high temperatures. The (V-I) curves show a positive curvature for
higher T and an ohmic behavior for low currents while for low T the
curves have a negative curvature. The onset of ohmic behavior
corresponds to a liquid phase and  the exponential decrease of the
voltage for decreasing values of the current is the signature of a
vortex glass phase~\cite{RHKoch89}.

In order to gain a  better understanding of the dynamic
response of this vortex structure, we use a scaling theory in the
critical region of a vortex-glass transition. In the scaling regime,
the relation between the electrical field $E$ and the current
density $J$ is given by
\begin{equation}
E|t|^{-\nu(z+2-D)}= F_{\pm}(|t|^{-(D-1)\nu}J),
\end{equation}
where $t=(T-T_G)/T_G$, $T_G$ is the temperature for the vortex glass
transition and $F_{\pm}$ are two universal functions for $t>0$ and
$t<0$~\cite{GBlatter94}. D is the dimensionality of the phase
transition, $z$ is the dynamical exponent and $\nu$ is the exponent
related to the divergence of the coherence length
($\xi\sim|T-T_G|^{-\nu}$). Using this scaling relation with $D=3$,
$T_G=3.15$ K, $\nu=3/8$ and $z=2.73$, the set of data from
Fig.~\ref{fig3}(a) ranging from 2.2 K to 3.8 K can be scaled to the
same universal functions as shown in Fig.~\ref{fig3}(c). This
universality lends strong support to the presence of a second order
phase transition. The values of the exponents $z$ and $\nu$ are
smaller but comparable to the exponents of the vortex glass
transition obtained for Y-Ba-Cu-O containing random point
defects~\cite{WJiang93} and for amorphous
superconductors~\cite{NCYeh93}. The difference in the exponents
gives us a transition with smaller correlation lengths and
relaxation rates.

In earlier experiments on embedded magnetic nanoparticles in SCs, an
increase of vortex pinning due to the presence of the magnetic
particles has been observed~\cite{NDRizzo96,THAlden66,ASnezhko05}.
In general, these pinning centers are extremely efficient and can be
originated by various physical
effects~\cite{GCarneiro04,MJVanbael01,ASnezhko05,APalau07}. In our
case, we have an even stronger constraint: the hybrid systems have
an intrinsic pinning, since the magnetic particles produce the
vortices and pin them to their original location. In this sense, the
difference in the scaling exponents should be related with the
different type of pinning centers created by magnetic nanoparticles.
We believe that an extension of the scaling theory addressing the
issue of magnetic pinning  is needed in order to fully understand this new spontaneous vortex solid (SVS).

Now we also can understand the observed sharp superconducting transition of the as-prepared sample shown in of Fig.~\ref{fig2} (a): due to the low deposition temperature of 40 K, the non-annealed sample has a large number of defects which results: (i) in a high resistivity of the normal state, (ii) a reduction of the $\xi$ ($\xi \propto l^{1/2}$) and (iii) in strong vortex pinning. Due to the later one the spontaneous vortex solid will remain up to the transition temperature $T_c$ resulting in a sharp transition at $T_c$.

For a further investigation of how the nanoparticles' magnetization
modifies the low temperature vortex state, we study the (V-I) curves
for B up to 0.01 T, which is much smaller than the field necessary
to align the magnetic particles (Fig.~\ref{fig1}(b)). The magnetic
field is applied parallel to the sample surface and we use the
following procedure: we first align the nanoparticles at room
temperature with a magnetic field $B=1$ T, in order to obtain a net
stray field of the Co particles ($B_{Co}$), i.e. $B_{Co}\neq 0$. Next, we field-cool the system and turn
off the field at a temperature above $T_c$ but far below the
blocking temperature of the particles.  We then apply the small
magnetic field in opposite directions at fixed temperature below
$T_c$. We find an asymmetric behavior of the (V-I) curves for
different polarities of the field (Fig.~\ref{fig3}(d)). This result
can be seen as a superposition of two different effects. The
relation of the total field (B$_T$) inside the sample, the external
field (B$_{ext}$) and the field of the Co clusters is
$B_{ T+}=B_{ext}+B_{Co}$ but with reversed external field the
relation becomes $B_{T-}= B_{ext}-B_{Co}$. This difference gives
rise to a very small shift in the critical current that depends on
the field direction. A similar result was observed in other systems
of nanomagnets inside a superconductor~\cite{APalau07}. A novel and
more pronounced effect occurs at small currents, when opposite
fields play the role of moving the system toward the phases above or
below the vortex glass transition.

\begin{figure}[h]
\includegraphics[width=0.8\columnwidth]{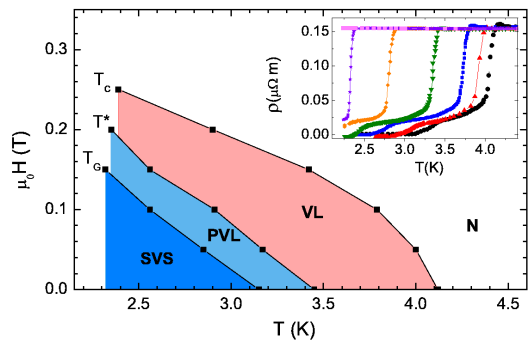}
\caption{(Color online) H-T phase diagram of the hybrid system containing Co
nanoparticles. SVS represents the spontaneous vortex solid. It is separated by critical line $T_G$ from a strongly pinned vortex liquid (PVL). The boundary to the unpinned vortex liquid
(VL) is $T^ *$, the characteristic temperature of the kink
feature seen in figure 2. The last critical curve is the main
superconducting transition at $T_c$ to the normal state (N). Inset: resistivity ($\rho$) versus temperature (T) for
various applied magnetic fields from 0 T to 0.3 T(right to left).
\label{fig4}}
\end{figure}

For applied fields higher than the coercive field of the sample, the
(V-I) curves do not depend on the polarity of the field. Therefore
we analyze the ($\rho$-T) (inset of Fig~\ref{fig4}) and (V-I) data
in the presence of an external magnetic field and construct an H-T
phase diagram. As before, the field is in-plane but now it is strong
enough to increase the net stray field produced by the Co particles
and the total field $B_{ T}$. As can be seen in the diagram of
Fig.~\ref{fig4}, it suppresses the superconductivity, shifting the
transitions ($T_G$, $T^*$ and $T_c$) to lower temperatures. Together
with the reduction of the critical temperatures, there is a shrink
in the SVS phase, showing that the intrinsic pinning is stronger
than the normal magnetic pinning. Finally, we can see that the lines $T_G$ and $T^*$ are
parallel to each other. It could indicate that the pinned liquid
state (PVL) is indeed a phase of partial movement, where the
vortices disentangle with reorientation and break of
loops~\cite{MVMilosevic08}. This non-linear effect could also be a consequence of
the inhomogeneous character of this superconducting phase and different nature
of pinning centers (magnetic and nonmagnetic) . This can constrain the vortex movement
and also give rise to regions with different pinning strengths.

In conclusion, our transport measurements show the signature of novel spontaneous vortex phases in superconducting films with
embedded magnetic particles.  At low temperatures, the spontaneous vortices are
organized in a disordered solid state (SVS), similar to a glass. For
increasing temperatures, the system undergoes a second order phase
transition to a pinned vortex liquid followed by a crossover that resembles
 vortex depinning. The dependence of the vortex solid state on the polarity
of the magnetic field demonstrates that ferromagnetic particles
inside superconductors can be used for specific forms of vortex
creation and manipulation.

This work was partially supported by CAPES/DAAD cooperation program
and the Brazilian agencies CNPq, FAPERJ (Cientistas do Nosso Estado
and  PRONEX) and L'Oreal Brazil. T. G. R and M. V. M would like to
thank ITS at UND for the hospitality. H. M acknowledges CAPES/DAAD
and PCI/CBPF for financial support.


\end{document}